\documentclass[journal]{IEEEtran}

\ifCLASSINFOpdf
\else
   \usepackage[dvips]{graphicx}
\fi
\usepackage{url}

\usepackage{graphicx}
\usepackage{amsmath,amssymb,amsfonts,amsthm}
\usepackage{hyperref}
\usepackage{tabularx}
\usepackage{verbatim}

\usepackage[caption=false]{subfig}

\usepackage{cite}
\usepackage[switch]{lineno}
\usepackage[named]{algo}
\usepackage[noend]{algpseudocode}
\usepackage{algorithm}

\usepackage{amssymb}
\usepackage{subfig}
\usepackage{float}
\usepackage{xcolor}

\newtheorem{theorem}{Theorem}

\newtheorem{proposition}[theorem]{Proposition}

\newtheorem{remark}{Remark}
\newcommand{\mbf}{\mathbf}

\begin{document}

\title{Coherent Source Enumeration with Compact ULAs}

\author{Dibakar Sil, Sunder Ram Krishnan and Kumar Vijay Mishra~\IEEEmembership{Senior Member,~IEEE}
\thanks{D. S. is an Independent Researcher, KA 560016, India. E-mail: dibakar.sil2@gmail.com.}
\thanks{S.~R.~K. is with the Amrita Vishwa Vidyapeetham, Amritapuri, KL 690525, India. E-mail: eeksunderram@gmail.com.}
\thanks{K. V. M. is with the University of Maryland, College Park, MD 20740, USA. E-mail: kvm@ieee.org.}}

\maketitle

\begin{abstract}
Source enumeration typically relies on subspace-based techniques that require accurate separation of signal and noise subspaces. However, prior works do not address coherent sources in small uniform linear arrays, where ambiguities arise in the spatial spectrum. We address this by decomposing the forward-backward smoothed covariance matrix into a sum of a rank-constrained Toeplitz matrix and a diagonal matrix with non-negative entries representing the signal and noise subspaces, respectively. The resulting non-convex optimization problem is solved by proposing  \textit{T}oeplitz \textit{a}pproach for \textit{r}ank-based tar\textit{g}et \textit{e}s\textit{t}imation (TARgEt) that employs the alternating direction method of multipliers. Numerical results on both synthetic and real-world datasets demonstrate the effectiveness and robustness of TARgEt over the state-of-the-art.
\end{abstract}


\IEEEpeerreviewmaketitle

\section{Introduction}
Source enumeration is a classic problem in array signal processing 
\cite{cozzens1994source,valaee2004information,garg2019subspace,ke2021accurate}. It requires estimating the number of signal sources impinging on an antenna array, wherein the observation vector, corrupted by additive noise, is modeled as a superposition of a finite number of source signals. This step is a prerequisite for algorithms such as MUSIC \cite{schmidt1986multiple} and ESPRIT \cite{roy2003esprit} for estimating parameters like  direction-of-arrival (DoA). 

Early approaches to source enumeration relied on hypothesis testing frameworks \cite{williams1990using,kritchman2009non}. To mitigate the subjectivity inherent in these methods, information-theoretic indicators \cite{wax1985detection} such as Akaike’s information criterion \cite{akaike1974new} and minimum description length \cite{rissanen1978modeling,rissanen2000mdl,schmidt2011consistency} were introduced. These techniques leveraged the spectral characteristics of the received signal sample covariance matrix in a uniform linear array (ULA), but were found to be most effective when a large number of observations are available \cite{huang2015bayesian}. Further, coherent sources result in performance degradation due to mismatch between the number of sources and sample covariance rank. 

For correlated/coherent sources, alternative strategies such as rank sequence analysis \cite{cozzens1994source}, spatial smoothing \cite{ma2002detection}, and signal subspace matching (SSM) \cite{wax2021detection} have been developed. A parametric model was developed in \cite{wax1989detection} using the idea that the number of sources can be more accurately estimated by benefiting from the knowledge of the estimated DoAs. Parametric models are better suited for enumerating correlated signals, especially those using matrix decomposition for source enumeration and DoA estimation. In this regard, \cite{wu2017toeplitz} proposed an optimization-based decomposition of the received signal covariance into the sum of a Toeplitz and diagonal matrix, but only considered fully uncorrelated sources. Without resorting to an optimization framework, \cite{zhang2019improved} handled coherent sources by exploiting half rows of the sample covariance with full information to reconstruct partial Toeplitz matrices. 

Source enumeration with coherent sources becomes particularly challenging in ULAs with very few sensors because closely-spaced sources tend to induce coherence in array response. Compact ULAs typically comprise around $[3K/2]$ antennas—the theoretical minimum \cite{pillai1989forward} needed to resolve $K$ sources, with $[\cdot]$ denoting the integer part. Such compact arrays commonly arise in low-frequency applications such as multi-robot systems \cite{sadler2024low}, underwater detection \cite{bekkerman2006target}, acoustics \cite{dokmanic2014hardware}, and in-cabin automotive radar \cite{fasraei2023ieee}. Among prior works, several large (i.e. $>>[3K/2]$) array source enumeration methods exist, addressing colored noise \cite{cardoso1990eigen, wu1995source}, low signal-to-noise ratio (SNR) \cite{chen1996detection}, and separate arrays \cite{lu2015source}. Recent approaches also include empirical model decomposition \cite{ge2022effective} and machine learning \cite{rogers2019estimating}. For large arrays, matrix decomposition methods often estimate DoAs by modeling the sample covariance as the sum of a low-rank positive semi-definite matrix and a diagonal matrix \cite{saunderson2012diagonal, qiu2016reduced, wang2025joint, malek2014doa}. However, a structured matrix decomposition-based optimization framework has not been investigated for compact ULA scenario with coherent sources.

In this paper, we address the enumeration for both fewer sensors and correlated sources. We propose decomposing the forward-backward (FB) smoothed covariance matrix (SCM) into a low-rank Toeplitz matrix and a diagonal matrix. This follows because the signal covariance in ULAs is approximately Toeplitz due to the array’s shift-invariant structure, while the noise covariance is diagonal under the assumption of spatially uncorrelated sensor noise. Previously, \cite{vallet2017performance} employed \textit{Toeplitz rectification}, which enforces a Toeplitz structure on the sample covariance matrix, to improve the performance of subspace methods under certain restrictive conditions (spatially uncorrelated sources with large number of antennas and snapshots). Next, we use spatially smoothed covariance in our optimization because it is known to approximate a noncoherent source model \cite{pillai1989forward}. Finally, a low-rank constraint on signal covariance arises because the number of sources is typically less than the number of sensors. 

Instead of subjective statistical model selection, our method adopts a geometric approach: the number of sources is simply the dimension of the signal subspace, which equals the column rank of the received signal covariance matrix. To recover both matrices, we introduce the \textit{T}oeplitz \textit{a}pproach for \textit{r}ank-based tar\textit{g}et \textit{e}s\textit{t}imation (TARgEt) algorithm that solves the resulting optimization problem using an alternating direction method of multipliers (ADMM). 
TARgEt considers coherent sources and includes the SCM in both the objective function and constraints. Although FB averaging is well-known, it is not straightforward to use it in a structured optimization framework for ULAs for improved performance.

Throughout this paper, we use bold lowercase and bold uppercase letters for vectors and matrices, respectively. The $i$-th element of a vector $\mathbf{x}$ is $x_i$; the $(i,j)$-th entry of a matrix $\mathbf{X}$ is $(\mathbf{X})_{i,j}$; $(\cdot)^T$ denotes the transpose of a vector/matrix; $\overline{\mbf{X}}$, $\mbf{X}^H$, $\|\mbf{X}\|_F$, $\textrm{rank}(\mbf{X})$, and $\|\mbf{X}\|_*$ denote, respectively, the conjugate, conjugate transpose, Frobenius norm, rank, and nuclear norm of matrix $\mbf{X}$. The notation $\mbf{X}\succeq\mbf{0}$ implies $\mbf{X}$ is positive semidefinite. 
The operator $\mathcal{P}_{\mbf{T}}(\cdot)$ projects a matrix onto the space of Toeplitz matrices:
$(\mathcal{P}_{\mbf{T}}(\mbf{X}))_{i, j} = \frac{1}{N - |i - j|} \sum_{k=1}^{N - |i - j|} (\mbf{X})_{i + k - 1, j + k - 1}$, where $\mbf{X}$ is of size $N$. The operator $\mathcal{D}_{\geq 0}(\cdot)$ extracts the diagonal elements of a matrix, enforces non-negativity, and returns a diagonal matrix. Also, $\mathcal{T}_K$ keeps the top $K$ elements in the diagonal of the input diagonal matrix of size $\geq K$, sets the rest to zero, and returns a diagonal matrix. 
The superscript in parenthesis $(\cdot)^{(i)}$ indicates value at $i$-th iteration.

\section{Signal Model}
Consider $K$ far-field narrowband sources from directions $\theta_i$, $i = 1, 2,\dots, K$, impinging on a ULA with $M$ isotropic sensors with interelement spacing of $d=\lambda/2$, where $\lambda$ is the wavelength. Denote the signal vector $\mbf{s}(t) = [s_1(t), \dots , s_K(t)]^T$, where $s_k(t)$ is the continuous-time signal of the $k$-th source, each with effective bandwidth $B_s$. 
Define the steering vector corresponding to the $k$-th source's angle-of-arrival (AoA) $\theta_k$ (its DoA being $ \sin \theta_k$) as  $\mathbf{a}(\theta_k) = \begin{bmatrix}
1, e^{\mathrm{j} \frac{2\pi}{\lambda} d \sin \theta_k}, 
\dots, e^{\mathrm{j} \frac{2\pi}{\lambda} (M-1) d \sin \theta_k}
\end{bmatrix}^T$. The $M\times 1$ received signal vector is a superposition of signals from all $K$ sources:\par\noindent\small 
\begin{align}
    \mbf{x}(t) = [x_1(t), \dots, x_M(t)]^T = \sum_{k=1}^K \mbf{a}(\theta_k)s_k(t) + \mbf{w}(t),
    \label{eq:conti_signal_model}
\end{align}\normalsize
where $\mbf{w}(t)$ is additive, spatio-temporally uncorrelated, zero-mean, non-white noise that is uncorrelated with the signal. 

The signal received in the $q$-th snapshot is acquired at sampling interval $T_s$ such that the discrete-time $M\times 1$ received signal vector is $\textbf{x}[q] = \textbf{x}(qT_s) = [x_1(qT_s),\dots,x_M(qT_s)]^T$. For all $Q$ snapshots, the ${M \times Q}$ sampled data matrix is 
\par\noindent\small
\begin{equation}
\mathbf{X} = [\mathbf{x}[1], \ldots, \mathbf{x}[Q]] = \mathbf{A}\mathbf{S} + \mathbf{W},
\label{eq:sig}
\end{equation}\normalsize
where 
$\mathbf{A} = [\mathbf{a}(\theta_1), \dots, \mathbf{a}(\theta_K)] \in \mathbb{C} ^{M \times K}$, $\mathbf{S} = [\mathbf{s}[1], \dots, \mathbf{s}[Q]] \in \mathbb{C}^{K \times Q}$, and $\mathbf{W} \in \mathbb{C} ^{M \times Q}$. 

A small number of elements $M$ or long wavelength $\lambda$ degrades ULA's angular resolution  $\Delta\phi = \frac{\lambda}{Md\cos(\theta)}$, limiting the ability to distinguish closely-spaced sources and increasing signal correlation \cite{rao2017mimo}. At low frequencies, the number of elements over a fixed aperture is inherently small. We therefore focus on estimating the number of (possibly coherent) sources $K$ under challenging conditions of low SNR and a sensor count $M$ near the theoretical resolution limit \cite{pillai1989forward,choi2002conditions}. 
The sample covariance matrix of the received signal is 
$(\mbf{R_X})_{j,k}=\frac{1}{Q}\displaystyle\sum_{i=1}^Q\left((\mbf{X})_{i,j}-\frac{1}{Q}\sum_{j=1}^Q(\mbf{X})_{i,j}\right)\left((\mbf{X})_{i,k}-\frac{1}{Q}\sum_{k=1}^Q(\mbf{X})_{i,j}\right)^H$. From the uncorrelatedness of signal and noise, we have 
$\mbf{R_X} =\mbf{AR_S A}^H+\mbf{R_W}$, where the source signal and noise covariances are, respectively, $\mbf{R_S}$ and a diagonal matrix $\mbf{R_W} \in \mathbb{C}^{M \times M}$. 
Denoting $\mbf{J} \in \mathbb{R}^{M \times M}$ as the exchange matrix (anti-diagonal identity), the spatially SCM, 
    $\mbf{R} = \frac{1}{2} (\mbf{R_X} + \mbf{J}\overline{\mbf{R}}_{\mbf{X}}\mbf{J})$,
is commonly employed to handle coherent sources \cite{evans1982application}. Denote the eigenvalues of $\mbf{R}$ sorted in descending order by $\sigma_1, \sigma_2,\dots,\sigma_M$.  

With a sufficient number of sensors, FB smoothing is known to enumerate coherent sources $K$ (\cite{choi2002conditions}). However, in practice, singular value decomposition (SVD) in the smoothed (auto)covariance matrix does not always efficiently resolve coherent sources \cite{zhang2019improved}. The eigenvalue spacing might confound the source enumeration because the most dominant eigenvalue $\sigma_1>>\sigma_2,\dots,\sigma_M$. For compact ULAs, singular value distributions are also ambiguous, which means it is difficult to clearly identify each singular value. 

%Drawing inspiration from prior works 
Following \cite{wu2017toeplitz,zhang2019improved}, we decompose the SCM as a sum of a Toeplitz matrix \(\mbf{L}\) and a diagonal matrix \(\mbf{D}\) representing the received signal and noise covariances, respectively. We account for a limited aperture via a rank constraint. This yields %the optimization problem
\par\noindent\small
\begin{equation}
\begin{aligned}
\underset{\mbf{L}, \mbf{D}}{\text{minimize}} \quad & \|(\mbf{L} + \mbf{D}) - \mbf{R}\|_F^2 + \eta (\text{rank}(\mbf{L})), \\
\text{subject to} \quad & (\mbf{L})_{i,j}=(\mbf{L})_{i+1,j+1}=\ell_{i-j}, \\
& \mbf{D}\succeq \mbf{0}, \quad(\mbf{D})_{i,j}=0,\,\forall i\neq j, \\
& \text{rank}(\mbf{L}) \leq K,
\end{aligned}
\label{eq: Opt}
\end{equation}\normalsize
where $\eta$ is a regularization parameter. A larger (smaller) $\eta$ favors a lower (higher) rank, risking source underestimation (noise overfitting). Thus, $\eta$ critically affects whether recovered rank matches $K$. Without the regularization term, this is a constrained least squares (CLS) problem:\par\noindent\small
\begin{align}
\underset{\mbf{L}, \mbf{D}}{\text{minimize}} \quad & \|(\mbf{L} + \mbf{D}) - \mbf{R}\|_F^2  \nonumber \\
\text{subject to} \quad & (\mbf{L})_{i,j}=(\mbf{L})_{i+1,j+1}=\ell_{i-j}, \nonumber\\
& \mbf{D}\succeq \mbf{0}, \quad(\mbf{D})_{i,j}=0,\,\forall i\neq j.\label{eq:noreg}
\end{align}\normalsize
Proposition \ref{lem:haha} below shows that it is possible to decompose the smoothed covariance into the sum of a Toeplitz matrix involving a full rank modified source signal covariance $\widetilde{\mbf{R}}_{\mbf{S}}$, and a diagonal noise covariance matrix.
\begin{proposition}
Consider the signal in \eqref{eq:sig} with $K$ (possibly) coherent sources. As $Q\rightarrow\infty$, the global optimum of the convex problem \eqref{eq:noreg} is $\mbf{L}^{\star}=\mathcal{P}_\mbf{T}(\mbf{A}\widetilde{\mbf{R}}_{\mbf{S}}\mbf{A}^H),$ 
and $\mbf{D}^{\star}=\mbf{R}_{\mbf{W}}$, where $\mbf{A}\widetilde{\mbf{R}}_{\mbf{S}}\mbf{A}^H$ is a size $M$ matrix of rank $K$.
\label{lem:haha}
\end{proposition}
\begin{IEEEproof}
Being a steering vector matrix of ULA, $\mbf{A}$ is an $M \times K$ Vandermonde matrix with distinct columns and, hence, rank $K$. As $Q\rightarrow\infty$, the sample covariance converges in probability to the population covariance. It follows from \cite[Section II]{choi2002conditions} that, with probability one, in the limit as $Q\rightarrow\infty$, $\mbf{R} \rightarrow \mbf{A}\widetilde{\mbf{R}}_{\mbf{S}}\mbf{A}^H+\mbf{R_W}$,
where $\widetilde{\mbf{R}}_{\mbf{S}}$ is a full rank size $K$ matrix. Then, the claim on the rank of $\mbf{A}\widetilde{\mbf{R}}_{\mbf{S}}\mbf{A}^H$ is trivial. Imposing the Toeplitz constraint projects $\mbf{A}\widetilde{\mbf{R}}_{\mbf{S}}\mbf{A}^H$ onto the space of Toeplitz matrices by operator $\mathcal{P}_\mbf{T}$. This completes the proof. 
\end{IEEEproof}
   \begin{remark}
        With the number of measurements $Q$ sufficiently large and when the modified source signal covariance $\widetilde{\mbf{R}}_{\mbf{S}}$ is strictly diagonally dominant\footnote{This follows from L\'{e}vy-Desplanques theorem \cite[p. 352]{horn2013matrix}, which states that strictly diagonally dominant matrices are nonsingular.}, Proposition \ref{lem:haha} implies that the error $\|(\mbf{L}^{\star} + \mbf{D}^{\star}) - \mbf{R}\|_F^2=\|(\mbf{I}-\mathcal{P}_\mbf{T})(\mbf{A}\widetilde{\mbf{R}}_{\mbf{S}}\mbf{A}^H)\|_F^2,$ is small. This follows from  $\mbf{A}\widetilde{\mbf{R}}_{\mbf{S}}\mbf{A}^H=\sum_{i=1}^M\sum_{k=1}^M (\widetilde{\mbf{R}}_{\mbf{S}})_{i,k}\mbf{a}_i\mbf{a}_k^H$ and the fact that, for a ULA, $\mbf{a}_i\mbf{a}_i^H$ is Toeplitz. This is effectively a Toeplitz rectification of the SCM, which improves the decorrelation of coherent sources and thereby enhances the performance of subspace-based methods.
    \end{remark}

Our goal is to recover $\mathbf{L}$ and $\mathbf{D}$, enforcing a low-rank constraint on $\mathbf{L}$ and selecting $\eta$ to yield an optimal $\mathbf{L}$ of rank $K$, thereby estimating the number of sources. The objective and constraints here are formulated differently from \cite{wu2017toeplitz}, but the key difference lies in our ability to treat the coherent source scenario. On the other hand, \cite{zhang2019improved} enhances DoA estimation via smoothed covariance and Toeplitz reconstruction using half rows of the sample covariance, but does not address source enumeration within an optimization framework.

\section{Recovery using TARgEt}
Rank constraint makes the optimization \eqref{eq: Opt} non-convex. Using nuclear norm $\|\mbf{L}\|_*$, its convex relaxation is \par\noindent\small
\begin{equation}
\label{eq:conv}
\begin{aligned}
\underset{\mbf{L}, \mbf{D}}{\text{minimize}} \quad & \|(\mbf{L} + \mbf{D}) - \mbf{R}\|_F^2  + \eta \|\mbf{L}\|_* \\
\text{subject to} \quad & (\mbf{L})_{i,j}=(\mbf{L})_{i+1,j+1}=\ell_{i-j}, \\
& \mbf{D}\succeq \mbf{0}, \quad(\mbf{D})_{i,j}=0,\,\forall i\neq j.
\end{aligned}
\end{equation}\normalsize
We propose to use ADMM to solve this problem because of its efficient practical implementations with large data and available theoretical convergence guarantees \cite{boyd2011distributed}. It is worth noting that previous related work on source enumeration used ADMM for similar reasons \cite{wu2017toeplitz}. 

Introducing auxiliary variables $\mbf{Z}$ and $\mbf{W}$ in \eqref{eq:conv} 
gives \noindent\par\small
\begin{align}
\label{eq:act}
\underset{\mbf{L}, \mbf{D}, \mbf{Z}, \mbf{W}}{\text{minimize}} \quad & \|(\mbf{L} + \mbf{D}) - \mbf{R}\|_F^2 + \eta \|\mbf{Z}\|_* \\
\text{subject to} \quad & \mbf{L} = \mbf{Z}, \; \mbf{D} = \mbf{W}, \nonumber \\
& (\mbf{L})_{i,j}=(\mbf{L})_{i+1,j+1}=\ell_{i-j}, \nonumber\\
& \mbf{W}\succeq \mbf{0}, \,(\mbf{W})_{i,j}=0,\,\forall i\neq j.\nonumber
\end{align}\normalsize
The augmented Lagrangian of \ref{eq:act} is $\mathcal{L}(\mbf{L}, \mbf{D}, \mbf{Z}, \mbf{W}, \mbf{Y}_1, \mbf{Y}_2, \mu)\nonumber\\
=  \|(\mbf{L} + \mbf{D}) - \mbf{R}\|_F^2 + \eta \cdot \|\mbf{Z}\|_* + \|\mbf{Y}_1^H (\mbf{L} - \mbf{Z})\|_* %\nonumber\\
 + \| \mbf{Y}_2^H (\mbf{D} - \mbf{W})\|_* + \frac{\mu}{2} \|\mbf{L} - \mbf{Z}\|_F^2 + \frac{\mu}{2} \|\mbf{D} - \mbf{W}\|_F^2,$
where $\mbf{Y}_1$ and $\mbf{Y}_2$ are dual variables and $\mu$ is the Lagrange multiplier. Singular value thresholding \cite{cai2010singular} is employed to solve the subproblem of nuclear norm minimization; the other subproblem is a CLS. The update equations for $\mathbf{L}$, $\mathbf{D}$,  $\mathbf{Z}$, and $\mathbf{W}$ are, respectively,\par\noindent\small
\begin{subequations}
\begin{align}
    \frac{\partial \mathcal{L}}{\partial \mathbf{L}} &= 2(\mathbf{L} + \mathbf{D} - \mathbf{R}) + \mathbf{Y}_1 + \mu(\mathbf{L} - \mathbf{Z}) = 0,\label{eq:Lderivative}\\
    \frac{\partial \mathcal{L}}{\partial \mathbf{D}} &= 2(\mathbf{L} + \mathbf{D} - \mathbf{R}) + \mathbf{Y}_2 + \mu(\mathbf{D} - \mathbf{W}) = 0,\label{eq:Dderivative}\\
    \frac{\partial \mathcal{L}}{\partial \mathbf{Z}} &= \eta \frac{\partial \|\mathbf{Z}\|_*}{\partial \mathbf{Z}} - \mathbf{Y}_1 + \mu(\mathbf{Z} - \mathbf{L}) = 0, \label{eq:Zderivative}\\
\textrm{and}\;     \frac{\partial \mathcal{L}}{\partial \mathbf{W}} &= -\mathbf{Y}_2 + \mu(\mathbf{W} - \mathbf{D}) = 0, \label{eq:Wderivative}
\end{align}
\end{subequations}
\normalsize
where we project the updated $\mbf{L}$ in \eqref{eq:Lderivative} ($\mbf{D}$ in \eqref{eq:Dderivative}) onto the space of Toeplitz (non-negative diagonal) matrices. Minimizing the nuclear norm in (\ref{eq:Zderivative}) encourages $\mathbf{Z}$ to be low rank and $\eta$ controls the strength of the low-rank constraint. The solution set $\mathbf{W}$ in (\ref{eq:Wderivative}) is the optimal diagonal approximation of $\mathbf{D}$. Algorithm~\ref{algo:TARgEt} summarizes the TARgEt method that outputs the number of sources as rank$(\widehat{\mbf{L}})$.
%--------------------------------------------------
\begin{algorithm}[H]
    \caption{TARgEt: Low-rank Toeplitz and diagonal matrix decomposition via ADMM}
    \begin{algorithmic}[1]
        \Statex \textbf{Input}: Augmented Lagrangian $\mathcal{L}$, threshold $\epsilon > 0$
        \Statex \textbf{Output}: Estimated number of sources $\text{rank}(\widehat{\mathbf{L}})$
        \State Initialize: $\mathbf{L}^{(0)}, \mathbf{D}^{(0)}, \mathbf{Z}^{(0)}, \mathbf{W}^{(0)}, \mathbf{Y}_1^{(0)}, \mathbf{Y}_2^{(0)}$ % Initial values for all variables
        \State Iteration $t \leftarrow 0$ % Iteration counter
        \While {$||\mathbf{L}^{(t)} - \mathbf{L}^{(t-1)}||_F \leq \epsilon$}
            \State $t \leftarrow t + 1$
            \State $\mathbf{L}^{(t)} \leftarrow \mathcal{P}_{\mbf{T}} \left( \frac{2(\mathbf{R} - \mathbf{D}^{(t-1)}) - \mathbf{Y}_1^{(t-1)} + \mu \mathbf{Z}^{(t-1)}}{2 + \mu} \right)$
            \State $\mathbf{D}^{(t)} \leftarrow \mathcal{D}_{\geq 0} \left( \frac{2(\mathbf{R} - \mathbf{L}^{(t)}) - \mathbf{Y}_2^{(t-1)} + \mu \mathbf{W}^{(t-1)}}{2 + \mu} \right)$
            \State $\mathbf{U}, \mathbf{\Sigma}, \mathbf{V} \leftarrow \text{SVD}(\mathbf{L}^{(t)} + \mathbf{Y}_1^{(t-1)} / \mu)$
            \State $\mathbf{Z}^{(t)} \leftarrow \mathbf{U}\, (\mathcal{T}_K(\mathcal{D}_{\geq 0}(\mathbf{\Sigma} - \eta/\mu))) \mathbf{V}^H$
            \State $\mathbf{W}^{(t)} \leftarrow \mathcal{D}_{\geq 0} (\mathbf{D}^{(t)} + \mathbf{Y}_2^{(t-1)} / \mu)$
            \State $\mathbf{Y}_1^{(t)} \leftarrow \mathbf{Y}_1^{(t-1)} + \mu (\mathbf{L}^{(t)} - \mathbf{Z}^{(t)})$
            \State $\mathbf{Y}_2^{(t)} \leftarrow \mathbf{Y}_2^{(t-1)} + \mu (\mathbf{D}^{(t)} - \mathbf{W}^{(t)})$
        \EndWhile
        \State $\widehat{\mathbf{L}} \leftarrow \mathbf{L}^{(t)}$
        \State \Return $\text{rank}(\widehat{\mathbf{L}})$
    \end{algorithmic}
    \label{algo:TARgEt}
\end{algorithm}
%--------------------------------------------------

\section{Numerical Experiments}
%--------------------------------------------------------
\begin{figure*}
    \centering    
    \includegraphics[width=1.0\textwidth]{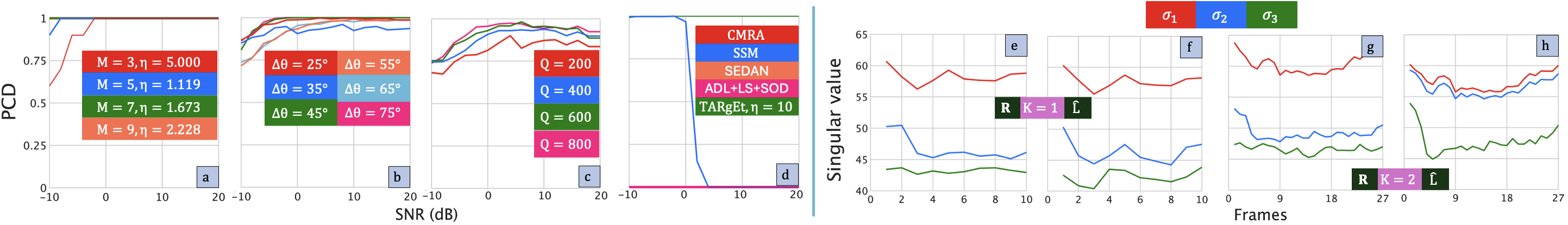} 
    \caption{Left panel: PCD versus SNR (a) using TARgEt for different numbers of antenna elements and fixed $K = 2$, $Q = 500$, $\Delta \theta = 70^\circ$, $\rho=0.8$  (b) using TARgEt for different angular separations and fixed $Q = 200$, $M = 3$, $K = 2$, $\rho=0.8$, and regularization parameter $\eta = 6$ (c) using TARgEt for different numbers of snapshots and fixed $M = 3$, $K = 2$, $\Delta \theta = 60^\circ$, $\rho=0.8$, and $\eta = 6$ (d) using TARgEt, SSM \cite{wax2021detection}, ADL+LS+SOD \cite{tian2024}, SEDAN \cite{muthukrishnan2024source}, and CMRA-ADMM \cite{wu2017toeplitz} for fixed $M = 3$, $K = 2$, $Q = 200$, $\Delta \theta = 70^\circ$, and $\rho=0.8$. Right panel: Singular values versus frames for (e) SCM $\mathbf{R}$ for $K = 1$, $M = 3$, $Q = 200$, and $\eta = 1$. (f) As in (e), but for Toeplitz-structured low rank matrix $\widehat{\mathbf{L}}$. (g) As in (e), but for $K=2$. (h) As in (e), but for $K=2$.}
    \label{fig:pcd_snr}
\end{figure*}
%--------------------------------------------------------
Our TARgEt method was evaluated using simulated and real-world ULA data. We consider compact ULAs with $M \in \{3, 5, 7, 9\}$ number of antennas. For quantifying effectiveness and performing comparisons, empirical probability of correct detection (PCD) over several independent Monte Carlo trials is employed as a metric. Only when the number of sources is exactly enumerated, do we mark it as a successful detection. Unless noted otherwise, throughout all experiments, we set $\lambda = 4$ cm and $d = 2$ cm. The number of independent simulation runs for each parameter setting was 200. In practice, noise is Gaussian in source enumeration setups and assumed so in our synthetic experiments.\\
\noindent\textbf{Synthetic Data:} Figure~\ref{fig:pcd_snr}a illustrates the PCD obtained via  TARgEt as a function of SNR with $K=2$, $Q=500$, and a fixed $\Delta\theta = 60^\circ$ for different $M$. While source enumeration accuracy generally improves under less noisy conditions, pre-tuning the regularization parameter $\eta$ is critical. Figure~\ref{fig:pcd_snr}b examines the algorithm's performance in resolving closely-spaced sources for fixed $M=3$, $K=2$, and $Q=500$. The PCD is plotted against SNR for various angular separations $\Delta \theta \in \{25^\circ, 35^\circ, 45^\circ, 55^\circ, 65^\circ, 75^\circ\}$. The performance at high SNRs shows the general trend of PCD to improve with increasing angular separation. However, this does not follow at low SNRs because the internal parameters of TARgEt require additional tuning for each angular separation. 
Figure~\ref{fig:pcd_snr}c depicts the impact of number of snapshots $Q$. Here, we set $M=3$, $K=2$, and $\Delta \theta = 60^\circ$. The PCD generally improves with increasing $Q$ %number of snapshots %($T = 200, 400, 600, 800, 1000$) 
across all tested SNR values. At SNR $=5$ dB, the PCD doubles from {\small $\sim$}$40$\% for $Q=200$ to {\small $\sim$}$80$\% for $Q=1000$ snapshots. In low-frequency applications such as Bluetooth and ultra-wideband, the data acquisition rate is of the order of milli-seconds ($1$-$5$ ms) \cite{lambrecht2025low} and the coherence time of a second, allowing large $Q$ values. Finally, Fig. 1d presents a performance comparison of TARgEt with several state-of-the-art algorithms, including SSM \cite{wax2021detection}, ADL (adaptive diagonal loading) + LS (linear shrinkage) + SOD (second-order difference) \cite{tian2024}, SEDAN (\textit{s}ource \textit{e}numeration using the \textit{d}istribution of \textit{an}gles) \cite{muthukrishnan2024source}, and CMRA-ADMM (a covariance matrix reconstruction method) \cite{wu2017toeplitz}. Since \cite{wu2017toeplitz} is not designed for coherent sources, we replace their covariance with its smoothed version. 

For $M=3$, Figure~\ref{fig:pcd_snr}d shows a significant performance improvement with TARgEt when the correlation coefficient of two signals is $\rho = \mathbb{E}[s_1(t)\overline{s_2}(t)]/(\sqrt{{\mathbb{E}[s_1(t)\overline{s_1}{(t)}]}} \sqrt{{\mathbb{E}[s_2(t)\overline{s_2}{(t)}]}})=0.8$. The TARgEt method exhibits consistently high PCD {\small$\sim$}$1.0$ across the $-10$ to $20$ dB SNR range. In contrast, ADL+LS+SOD and SEDAN demonstrate poor empirical results in this setting. When practical conditions, such as a small number of sensors relative to the sources, do not align with the random matrix theory-based asymptotic premises made in ADL+LS+SOD, the algorithm's ability to accurately differentiate between signal and noise subspaces is compromised, leading to degraded performance. For SEDAN, this is likely due to its reliance on angular distribution differences, which tend to be unstable, especially in scenarios with small sensor arrays where geometric diversity is inherently limited. 
While SSM performs well at very low SNRs ($<0$ dB), its PCD drops sharply near 0 dB and remains low at higher SNRs, indicating significant subspace distortion due to source correlation and limited array aperture. CMRA-ADMM exhibits consistently poor PCD across all SNRs. In contrast, TARgEt achieves superior performance because of its effective decomposition of the covariance matrix, which mitigates the effects of correlation. SSM is more vulnerable to subspace degradation, while CMRA-ADMM’s rigid Toeplitz enforcement under constrained settings limits its adaptability.\\
\noindent\textbf{Real Data:}  
We collected real-world data using a 3-element ULA with $\lambda / d = 2$. In the first scenario,  only one stationary source with radar cross-section (RCS) $0$ dBsm was located $2$ m from the ULA, with the target positioned at AoA $0^\circ$ with respect to the array's boresight. In the second case, two stationary sources, each with RCS of $0$ dBsm, were located $2$m from the ULA. The targets were positioned at AoAs  $-30^\circ$ and $50^\circ$ with respect to the array's boresight. Define a quantitative measure of the eigenvalue spacing of a matrix $\mbf{C}$ as $r(\mbf{C})=(\overline{\sigma}_1(\mbf{C})-\overline{\sigma}_2(\mbf{C}))/(\overline{\sigma}_2(\mbf{C})-\overline{\sigma}_3(\mbf{C}))$, where $\overline{\sigma}_i(\mbf{C})$ denotes the average of the $i$-th eigenvalue across frames, where a frame comprises 200 snapshots. 
For only one source in the boresight, the SVD of estimated Toeplitz matrix $\widehat{\mbf{L}}$ (and $\mbf{R}$) exhibited a clearly dominant singular value, with the other two singular values considerably smaller and comparable in magnitude, indicative of the noise subspace. This supports the presence of a rank-1 signal subspace corresponding to the single source. We obtained $r(\mbf{R})=3.35$ and $r(\widehat{\mbf{L}})=2.90$ averaged over 9 frames. Note that $r(\widehat{\mbf{L}})$ remains tightly bounded within $\pm 0.03$ of $2.90$ as $\eta$ ranges from $1$ to $5$. Increasing $\eta$ beyond $5$ resulted in $r(\widehat{\mbf{L}})$ degrading by $62\%$. 
For two distinct sources with an approximate angular separation of $80^\circ$, the SVD of $\widehat{\mbf{L}}$ consistently revealed two dominant singular values, significantly larger than the other singular values. This aligns with the expectation of a rank-$2$ signal subspace corresponding to two sources. From the average values across 27 frames -- specifically $r(\mbf{R})=4.30$ and $r(\widehat{\mbf{L}})=0.13$ (with $r(\widehat{\mbf{L}}) \in (0.11,0.15)$ when $\eta\in[1,5]$) -- we infer that while the spacing of the eigenvalues of $\mbf{R}$ suggests the presence of a single source, $\widehat{\mbf{L}}$ reliably indicates two. The quantitative measure $r$ suggests that both $\mbf{R}$ and $\widehat{\mbf{L}}$ effectively solve the single-source scenario. However, taken together with the values of $r$ for the two-source scenario, it may be concluded that only $\widehat{\mathbf{L}}$ accurately reflects the expected rank-2 subspace structure; the result obtained with $\mbf{R}$ shows that it cannot separate the one vs. two-source cases. Moreover, the insensitivity of $r(\widehat{\mbf{L}})$ to mild changes in $\eta$ confirms that the algorithm remains robust across a range of $\eta$ values.
Figure~\ref{fig:pcd_snr}e-h show plots of eigenvalues with respect to the number of frames, demonstrating the potential of the proposed method to correctly enumerate active sources by analyzing the rank of the estimated matrix $\widehat{\mbf{L}}$. 
\section{Summary}
Classical source enumeration has mainly focused on large arrays. For small ULAs, we express the FB smoothed covariance in terms of a modified, full-rank source covariance, leading to an optimization that decomposes it into a rank-constrained Toeplitz and a diagonal matrix. Similar low-rank decompositions appear in space-time adaptive processing for airborne radars \cite{sen2015low}, where the number of antennas is significantly lower than the synthesized aperture \cite{vouras2023overview}, and in the study of elliptopes \cite{sadler2024low}, with connections to recent machine learning theory \cite{brockmeier2017quantifying}.

\clearpage
\bibliographystyle{IEEEtran}
\bibliography{main}

\begin{thebibliography}{10}
\providecommand{\url}[1]{#1}
\csname url@samestyle\endcsname
\providecommand{\newblock}{\relax}
\providecommand{\bibinfo}[2]{#2}
\providecommand{\BIBentrySTDinterwordspacing}{\spaceskip=0pt\relax}
\providecommand{\BIBentryALTinterwordstretchfactor}{4}
\providecommand{\BIBentryALTinterwordspacing}{\spaceskip=\fontdimen2\font plus
\BIBentryALTinterwordstretchfactor\fontdimen3\font minus \fontdimen4\font\relax}
\providecommand{\BIBforeignlanguage}[2]{{%
\expandafter\ifx\csname l@#1\endcsname\relax
\typeout{** WARNING: IEEEtran.bst: No hyphenation pattern has been}%
\typeout{** loaded for the language `#1'. Using the pattern for}%
\typeout{** the default language instead.}%
\else
\language=\csname l@#1\endcsname
\fi
#2}}
\providecommand{\BIBdecl}{\relax}
\BIBdecl

\bibitem{cozzens1994source}
J.~H. Cozzens and M.~J. Sousa, ``Source enumeration in a correlated signal environment,'' \emph{IEEE Transactions on Signal Processing}, vol.~42, no.~2, pp. 304--317, 1994.

\bibitem{valaee2004information}
S.~Valaee and P.~Kabal, ``An information theoretic approach to source enumeration in array signal processing,'' \emph{IEEE Transactions on Signal Processing}, vol.~52, no.~5, pp. 1171--1178, 2004.

\bibitem{garg2019subspace}
V.~Garg, I.~Santamaria, D.~Ramirez, and L.~L. Scharf, ``Subspace averaging and order determination for source enumeration,'' \emph{IEEE Transactions on Signal Processing}, vol.~67, no.~11, pp. 3028--3041, 2019.

\bibitem{ke2021accurate}
X.~Ke, Y.~Zhao, and L.~Huang, ``On accurate source enumeration: A new {B}ayesian information criterion,'' \emph{IEEE Transactions on Signal Processing}, vol.~69, pp. 1012--1027, 2021.

\bibitem{schmidt1986multiple}
R.~Schmidt, ``Multiple emitter location and signal parameter estimation,'' \emph{IEEE Transactions on Antennas and Propagation}, vol.~34, no.~3, pp. 276--280, 1986.

\bibitem{roy2003esprit}
R.~Roy, A.~Paulraj, and T.~Kailath, ``{ESPRIT}--a subspace rotation approach to estimation of parameters of cisoids in noise,'' \emph{IEEE Transactions on Acoustics, Speech, and Signal Processing}, vol.~34, no.~5, pp. 1340--1342, 2003.

\bibitem{williams1990using}
D.~B. Williams and D.~H. Johnson, ``Using the sphericity test for source detection with narrow-band passive arrays,'' \emph{IEEE Transactions on Acoustics, Speech, and Signal Processing}, vol.~38, no.~11, pp. 2008--2014, 1990.

\bibitem{kritchman2009non}
S.~Kritchman and B.~Nadler, ``Non-parametric detection of the number of signals: {H}ypothesis testing and random matrix theory,'' \emph{IEEE Transactions on Signal Processing}, vol.~57, no.~10, pp. 3930--3941, 2009.

\bibitem{wax1985detection}
M.~Wax and T.~Kailath, ``Detection of signals by information theoretic criteria,'' \emph{IEEE Transactions on Acoustics, Speech, and Signal Processing}, vol.~33, no.~2, pp. 387--392, 1985.

\bibitem{akaike1974new}
H.~Akaike, ``A new look at the statistical model identification,'' \emph{IEEE Transactions on Automatic Control}, vol.~19, no.~6, pp. 716--723, 1974.

\bibitem{rissanen1978modeling}
J.~Rissanen, ``Modeling by shortest data description,'' \emph{Automatica}, vol.~14, no.~5, pp. 465--471, 1978.

\bibitem{rissanen2000mdl}
------, ``{MDL} denoising,'' \emph{IEEE Transactions on Information Theory}, vol.~46, no.~7, pp. 2537--2543, 2000.

\bibitem{schmidt2011consistency}
D.~F. Schmidt and E.~Makalic, ``The consistency of {MDL} for linear regression models with increasing signal-to-noise ratio,'' \emph{IEEE Transactions on Signal Processing}, vol.~60, no.~3, pp. 1508--1510, 2011.

\bibitem{huang2015bayesian}
L.~Huang, Y.~Xiao, K.~Liu, H.~C. So, and J.-K. Zhang, ``Bayesian information criterion for source enumeration in large-scale adaptive antenna array,'' \emph{IEEE Transactions on Vehicular Technology}, vol.~65, no.~5, pp. 3018--3032, 2015.

\bibitem{ma2002detection}
C.-W. Ma and C.-C. Teng, ``Detection of coherent signals using weighted subspace smoothing,'' \emph{IEEE Transactions on Antennas and Propagation}, vol.~44, no.~2, pp. 179--187, 2002.

\bibitem{wax2021detection}
M.~Wax and A.~Adler, ``Detection of the number of signals by signal subspace matching,'' \emph{IEEE Transactions on Signal Processing}, vol.~69, pp. 973--985, 2021.

\bibitem{wax1989detection}
M.~Wax and I.~Ziskind, ``Detection of the number of coherent signals by the {MDL} principle,'' \emph{IEEE Transactions on Acoustics, Speech, and Signal Processing}, vol.~37, no.~8, pp. 1190--1196, 1989.

\bibitem{wu2017toeplitz}
X.~Wu, W.-P. Zhu, and J.~Yan, ``A {T}oeplitz covariance matrix reconstruction approach for direction-of-arrival estimation,'' \emph{IEEE Transactions on Vehicular Technology}, vol.~66, no.~9, pp. 8223--8237, 2017.

\bibitem{zhang2019improved}
W.~Zhang, Y.~Han, M.~Jin, and X.-S. Li, ``An improved {ESPRIT}-like algorithm for coherent signals {DOA} estimation,'' \emph{IEEE Communications Letters}, vol.~24, no.~2, pp. 339--343, 2019.

\bibitem{pillai1989forward}
S.~U. Pillai and B.~H. Kwon, ``Forward/backward spatial smoothing techniques for coherent signal identification,'' \emph{IEEE Transactions on Acoustics, Speech, and Signal Processing}, vol.~37, no.~1, pp. 8--15, 1989.

\bibitem{sadler2024low}
B.~M. Sadler, F.~T. Dagefu, J.~N. Twigg, G.~Verma, P.~Spasojevic, R.~J. Kozick, and J.~Kong, ``Low frequency multi-robot networking,'' \emph{IEEE Access}, vol.~12, pp. 21\,954--21\,984, 2024.

\bibitem{bekkerman2006target}
I.~Bekkerman and J.~Tabrikian, ``Target detection and localization using {MIMO} radars and sonars,'' \emph{IEEE Transactions on Signal Processing}, vol.~54, no.~10, pp. 3873--3883, 2006.

\bibitem{dokmanic2014hardware}
I.~Dokmani{\'c} and I.~Tashev, ``Hardware and algorithms for ultrasonic depth imaging,'' in \emph{IEEE International Conference on Acoustics, Speech and Signal Processing}, 2014, pp. 6702--6706.

\bibitem{fasraei2023ieee}
A.~Farsaei, B.~Meyer, A.~Sheikh, M.~El~Soussi, P.~Zhang, G.~K. Ramachandra, J.~Govers, and M.~Hijdra, ``An {IEEE} 802.15.4z-compliant {IR-UWB} radar system for in-cabin monitoring,'' in \emph{IEEE Annual International Symposium on Personal, Indoor and Mobile Radio Communications}, 2023, pp. 1--5.

\bibitem{cardoso1990eigen}
J.-F. Cardoso, ``Eigen-structure of the fourth-order cumulant tensor with application to the blind source separation problem,'' in \emph{IEEE International Conference on Acoustics, Speech, and Signal Processing}, 1990, pp. 2655--2658.

\bibitem{wu1995source}
H.-T. Wu, J.-F. Yang, and F.-K. Chen, ``Source number estimators using transformed {G}erschgorin radii,'' \emph{IEEE Transactions on Signal Processing}, vol.~43, no.~6, pp. 1325--1333, 1995.

\bibitem{chen1996detection}
W.~Chen, J.~Reilly, and K.~Wong, ``Detection of the number of signals in noise with banded covariance matrices,'' \emph{IEE Proceedings - Radar, Sonar and Navigation}, vol. 143, no.~5, pp. 289--294, 1996.

\bibitem{lu2015source}
Z.~Lu and A.~M. Zoubir, ``Source enumeration in array processing using a two-step test,'' \emph{IEEE Transactions on Signal Processing}, vol.~63, no.~10, pp. 2718--2727, 2015.

\bibitem{ge2022effective}
S.~Ge, S.~N. B.~M. Rum, H.~Ibrahim, E.~Marsilah, and T.~Perumal, ``An effective source number enumeration approach based on {SEMD},'' \emph{IEEE Access}, vol.~10, pp. 96\,066--96\,078, 2022.

\bibitem{rogers2019estimating}
J.~Rogers, J.~E. Ball, and A.~C. Gurbuz, ``Estimating the number of sources via deep learning,'' in \emph{IEEE Radar Conference}, 2019, pp. 1--5.

\bibitem{saunderson2012diagonal}
J.~Saunderson, V.~Chandrasekaran, P.~A. Parrilo, and A.~S. Willsky, ``Diagonal and low-rank matrix decompositions, correlation matrices, and ellipsoid fitting,'' \emph{SIAM Journal on Matrix Analysis and Applications}, vol.~33, no.~4, pp. 1395--1416, 2012.

\bibitem{qiu2016reduced}
L.~Qiu, Y.~Cai, R.~C. de~Lamare, and M.~Zhao, ``Reduced-rank {DOA} estimation algorithms based on alternating low-rank decomposition,'' \emph{IEEE Signal Processing Letters}, vol.~23, no.~5, pp. 565--569, 2016.

\bibitem{wang2025joint}
X.-Y. Wang, X.-P. Li, H.~Huang, and H.~C. So, ``Joint {DOA} estimation and distorted sensor detection,'' \emph{IEEE Transactions on Aerospace and Electronic Systems}, 2025, in press.

\bibitem{malek2014doa}
M.~Malek-Mohammadi, M.~Jansson, A.~Owrang, A.~Koochakzadeh, and M.~Babaie-Zadeh, ``{DOA} estimation in partially correlated noise using low-rank/sparse matrix decomposition,'' in \emph{IEEE Sensor Array and Multichannel Signal Processing Workshop}, 2014, pp. 373--376.

\bibitem{vallet2017performance}
P.~Vallet and P.~Loubaton, ``On the performance of {MUSIC} with {T}oeplitz rectification in the context of large arrays,'' \emph{IEEE Transactions on Signal Processing}, vol.~65, no.~22, pp. 5848--5859, 2017.

\bibitem{rao2017mimo}
S.~Rao, ``{MIMO} radar,'' Texas Instruments, Tech. Rep. SWRA554A, May 2017.

\bibitem{choi2002conditions}
Y.-H. Choi, ``On conditions for the rank restoration in forward/backward spatial smoothing,'' \emph{IEEE Transactions on Signal Processing}, vol.~50, no.~11, pp. 2900--2901, 2002.

\bibitem{evans1982application}
J.~E. Evans, D.~Sun, and J.~Johnson, ``Application of advanced signal processing techniques to angle of arrival estimation in {ATC} navigation and surveillance systems,'' MIT Lincon Laboratory, USA, Tech. Rep. TR-582, FAA-RD-82-42, 1982.

\bibitem{horn2013matrix}
R.~A. Horn and C.~R. Johnson, \emph{Matrix analysis}, 2nd~ed.\hskip 1em plus 0.5em minus 0.4em\relax Cambridge University Press, 2013.

\bibitem{boyd2011distributed}
S.~Boyd, N.~Parikh, E.~Chu, B.~Peleato, J.~Eckstein \emph{et~al.}, \emph{Distributed optimization and statistical learning via the alternating direction method of multipliers}, ser. Foundations and Trends{\textregistered} in Machine learning.\hskip 1em plus 0.5em minus 0.4em\relax Now Publishers, Inc., 2011, vol.~3, no.~1.

\bibitem{cai2010singular}
J.-F. Cai, E.~J. Cand{\`e}s, and Z.~Shen, ``A singular value thresholding algorithm for matrix completion,'' \emph{SIAM Journal on optimization}, vol.~20, no.~4, pp. 1956--1982, 2010.

\bibitem{tian2024}
Y.~Tian, Z.~Zhang, W.~Liu, H.~Chen, and G.~Wang, ``Source enumeration utilizing adaptive diagonal loading and linear shrinkage coefficients,'' \emph{IEEE Transactions on Signal Processing}, vol.~72, pp. 2073--2086, 2024.

\bibitem{muthukrishnan2024source}
G.~Muthukrishnan, S.~Shanmugam, and S.~Kalyani, ``Source enumeration using the distribution of angles: {A} robust and parameter-free approach,'' \emph{arXiv preprint arXiv:2409.06563}, 2024.

\bibitem{lambrecht2025low}
A.~Lambrecht, S.~Luchie, J.~Fontaine, B.~Van~Herbruggen, A.~Shahid, and E.~De~Poorter, ``Low-cost embedded breathing rate determination using 802.15.4z {IR-UWB} hardware for remote healthcare,'' \emph{arXiv preprint arXiv:2504.03772}, 2025.

\bibitem{sen2015low}
S.~Sen, ``Low-rank matrix decomposition and spatio-temporal sparse recovery for {STAP} radar,'' \emph{IEEE Journal of Selected Topics in Signal Processing}, vol.~9, no.~8, pp. 1510--1523, 2015.

\bibitem{vouras2023overview}
P.~Vouras, K.~V. Mishra, A.~Artusio-Glimpse, S.~Pinilla, A.~Xenaki, D.~W. Griffith, and K.~Egiazarian, ``An overview of advances in signal processing techniques for classical and quantum wideband synthetic apertures,'' \emph{IEEE Journal of Selected Topics in Signal Processing}, vol.~17, no.~2, pp. 317--369, 2023.

\bibitem{brockmeier2017quantifying}
A.~J. Brockmeier, T.~Mu, S.~Ananiadou, and J.~Y. Goulermas, ``Quantifying the informativeness of similarity measurements,'' \emph{Journal of Machine Learning Research}, vol.~18, no.~76, pp. 1--61, 2017.

\end{thebibliography}

\end{document}